\documentclass[preprintnumbers, floatfix, preprintnumbers, letterpaper, superscriptaddress,nofootinbib,aps]{revtex4-2}
\pdfoutput=1
\usepackage{graphicx}
\usepackage{microtype}
\usepackage{amsmath}
\usepackage{amssymb}
\usepackage{subfigure}
\usepackage{hyperref}
\usepackage{url}
\usepackage{xcolor}
\usepackage{color}
\usepackage{mathrsfs}
\usepackage{calrsfs}
\usepackage{amsfonts}
\usepackage{latexsym}
\usepackage{ragged2e}
\usepackage{epstopdf}
\usepackage{textcomp}
\usepackage{phaistos}
\usepackage[utf8]{inputenc}
\usepackage{ulem}
\usepackage{float}
\usepackage{natbib}

\makeatletter
\renewcommand\@makefnmark{\hbox{\@textsuperscript{\normalfont\color{purple}\@thefnmark}}}
\renewcommand\@makefntext[1]{%
  \parindent 1em\noindent
            \hb@xt@1.8em{%
                \hss\@textsuperscript{\normalfont\@thefnmark}}#1}
\makeatother

\usepackage{caption}
\DeclareCaptionJustification{justified}{\leftskip=0pt \rightskip=0pt \parfillskip=0pt plus 1fil}
\definecolor{vividviolet}{rgb}{0.62, 0.0, 1.0}
\definecolor{amaranth}{rgb}{0.9, 0.17, 0.31}
\definecolor{palatinateblue}{rgb}{0.15, 0.23, 0.89}
\definecolor{brightpink}{rgb}{1.0, 0.0, 0.5}
\definecolor{cornflowerblue}{rgb}{0.39, 0.58, 0.93}
\definecolor{deepcarminepink}{rgb}{0.94, 0.19, 0.22}
\definecolor{radicalred}{rgb}{1.0, 0.21, 0.37}

\hypersetup{ linktoc=all,
    colorlinks, linkcolor={palatinateblue},
    citecolor={brightpink}, urlcolor={amaranth}
}

\graphicspath{{Images/}}

\def\sideremark#1{\ifvmode\leavevmode\fi\vadjust{\vbox to0pt{\vss
 \hbox to 0pt{\hskip\hsize\hskip1em
 \vbox{\hsize1.5cm\tiny\raggedright\pretolerance10000
 \noindent #1\hfill}\hss}\vbox to8pt{\vfil}\vss}}}%
\begin{document}

\title{The Interior of the Scalar Hairy Black Hole with Inverted Higgs Potential}

\author{Xiao Yan \surname{Chew}}
\email{xiao.yan.chew@just.edu.cn}
\affiliation{School of Science, Jiangsu University of Science and Technology, 212100, Zhenjiang, China.}

\author{Kok-Geng \surname{Lim}}
\email{K.G.Lim@soton.ac.uk}
\affiliation{Smart Manufacturing and Systems Research Group, University of Southampton Malaysia, 79100 Iskandar Puteri, Malaysia.}

\author{Dong-han \surname{Yeom}}
\email{innocent.yeom@gmail.com}
\affiliation{Department of Physics Education, Pusan National University, Busan 46241, Republic of Korea}
\affiliation{Research Center for Dielectric and Advanced Matter Physics, Pusan National University, Busan 46241, Republic of Korea}
\affiliation{Leung Center for Cosmology and Particle Astrophysics, National Taiwan University, Taipei 10617, Taiwan }
\affiliation{Department of Physics and Astronomy, University of Waterloo, Waterloo, ON N2L 3G1, Canada}

\begin{abstract}
We investigate the interior structure of asymptotically flat hairy black holes (HBHs) arising in the Einstein–Klein–Gordon theory with nonpositive-definite scalar potentials, where nontrivial scalar hair exists at the event horizon. While exterior properties, including shadow imaging for HBHs supported by an inverted Higgs-like potential have been extensively investigated, their interior structure remains largely unexplored. In many gravitational theories, backreaction of classical fields can significantly eliminate the Cauchy horizon, which is known to be highly unstable due to the mass inflation effect, raising important questions regarding the validity of the Strong Cosmic Censorship conjecture. These considerations motivate us to examine the interior structure of HBHs by numerically integrating the field equations inward from the outer horizon. We find that the scalar field and the metric functions increase monotonically inside the horizon and diverge as $r \rightarrow 0$. The Ricci and Kretschmann scalars also diverge at $r=0$, confirming the presence of a genuine curvature singularity. No additional root of the metric function is observed, indicating the absence of a Cauchy horizon in the electrically neutral HBHs considered here. Furthermore, the weak energy condition is violated throughout the interior region, and the degree of violation becomes more pronounced as the scalar field at the horizon increases. These results provide new insight into the global structure of HBHs and their implications for cosmic censorship.
\end{abstract}

\maketitle

\section{Introduction}

Black holes are among the most compelling predictions of General Relativity (GR). Their existence is now supported by robust observational evidence. In particular, the detection of gravitational waves from merger of binary black holes by the LIGO–VIRGO–KAGRA collaboration \cite{LIGOScientific:2016aoc,KAGRA:2013rdx,Cahillane:2022pqm} and the direct imaging of black hole shadows by the Event Horizon Telescope \cite{EventHorizonTelescope:2021dqv,EventHorizonTelescope:2022wkp,EventHorizonTelescope:2022xqj}. These observations provide precise confirmation of the external properties of black holes, but their interior structure behind the event horizon still remains inaccessible to direct measurement. Therefore, exploring this hidden region becomes an important theoretical interest to understand the global structure of black hole spacetime.

In general, the interior of a black hole, beyond its event horizon, contains a singularity, where the curvature of spacetime diverges, requiring new physics for any robust predictions in this extreme regime. However, another important feature of black hole interiors is the Cauchy horizon, which plays a fundamental role in the predictability of GR. We shall begin with a well-known example, which is the Reissner-Nordstr\"om black hole (RNBH), a static charged black hole that possesses both an outer event horizon and an inner Cauchy horizon. However, the Cauchy horizon is highly unstable due to infinite blue-shift instability, where small perturbations can lead to mass inflation, causing the energy density to grow exponentially \cite{Simpson:1973ua,Poisson:1990eh,Ori:1991zz,Bonanno:1994ma,Ong:2020xwv}. Consequently, the elimination of the Cauchy horizon could stabilize black hole interiors and reinforce the validity of Strong Cosmic Censorship Conjecture (SCCC), asserting that the spacetime can be completely deterministic when the unstable Cauchy horizon breaks down. 

A promising approach to investigate this phenomenon lies in the study of hairy black holes (HBHs), where one typically integrates the field equations inward from the event horizon, as the boundary conditions at the event horizon are already determined by the exterior solutions. Nontrivial classical fields at the event horizon not only allow HBHs to be described by extra physical quantities other than their mass, electric charge, and angular momentum, but also significantly impact their interior structures. In particular, the systematic investigations of HBHs in non-Abelian theories have shown that although the Cauchy horizons generally do not exist \cite{Donets:1998xd,Donets:1996ja,Sarbach:1997us,Breitenlohner:1997hm,Galtsov:1997ub}, some interesting phenomena could occur inside these HBHs. For instance, the exterior solutions of HBHs in the Einstein-Yang-Mills theory are shown to be unstable under linear perturbation, but the interior structure remains remarkably stable \cite{Donets:1998xd}. Moreover, its mass function oscillates infinitely near the singularity, while the Yang-Mills gauge field remains nearly constant \cite{Donets:1996ja}. The introduction of additional fields, such as a dilaton field in Einstein-Yang-Mills-dilaton (EYMd) theory and a Higgs field in Einstein-Yang-Mills-Higgs (EYMH) theory does not significantly change the exterior structure of HBHs, but can dramatically reshape their interior. For instance: In EYMd theory, the number of oscillations in the interior metric depends on the dilatonic coupling constant $\gamma$, with more oscillations occurring for smaller values of $\gamma$ \cite{Sarbach:1997us}; In EYMH theory, the introduction of a Higgs field eliminates the oscillatory behavior, resulting in a qualitatively different interior structure \cite{Breitenlohner:1997hm,Galtsov:1997ub}.

The absence of the Cauchy horizon is also particularly significant in the Anti-de 
Sitter/Conformal Field Theory (AdS/CFT) correspondence, as it maintains the robustness of the duality between gravitational theories in the bulk and CFT at the boundary. A notable study explored the dynamics of a charged HBH in AdS space in the context of holographic superconductors \cite{Hartnoll:2020rwq,Hartnoll:2020fhc}. Their findings revealed that when the temperature falls below a critical value, a charged HBH emerges from a Reissner-Nordstr\"om-AdS black hole via scalar hair formation. They found that the HBH interior inevitably collapses into a spacelike singularity, preserving the SCCC, since the Cauchy horizon doesn’t exist. Most strikingly, their study identified four distinct dynamical epochs approaching the singularity: collapse of the Einstein-Rosen bridge; rapid oscillations of Josephson-like; Kasner-like epoch; Kasner inversion transitions. Other investigations into static charged HBHs for holographic superconductor with various scalar potentials, spacetime dimensions and in 
different gravitational theories also exhibit these four stages of interior dynamics and found no evidence of the Cauchy horizon, reinforcing the phenomena that its absence is a generic feature of HBHs \cite{Wang:2020nkd,Cai:2021obq,Henneaux:2022ijt,An:2022lvo,Hartnoll:2022rdv,Gao:2023zbd,Caceres:2023zhl,Gao:2023rqc,Xu:2023fad,Cai:2023igv,Arean:2024pzo,Carballo:2024hem,Cai:2024ltu,Caceres:2024edr,Zhang:2025hkb,Zhang:2025tsa,Xu:2025edz,Li:2023tfa,Zhao:2025odj,Policastro:2025odh,Faraoni:2025ufi}.

Furthermore, both analytical and numerical analyses on a class of static, asymptotically flat and charged HBHs with spherical and planar horizons \cite{Cai:2020wrp} demonstrated similar conclusion to Ref. \cite{Hartnoll:2020fhc}, but charged HBHs with hyperbolic horizon can admit at most one inner horizon. This reveals the impact of horizon topology on the existence of Cauchy horizon. Similar findings on non-existence of Cauchy horizon can also be extended to other classes of asymptotically flat HBHs \cite{Kleihaus:2007cf,Brihaye:2016vkv,Grandi:2021ajl,An:2021plu,Brihaye:2021mqk,Dias:2021afz,Yang:2021civ,Devecioglu:2023hmn,Shao:2025apr,LiLi:2025qgo,Duan:2026qhj,Xiong:2026npi}. Besides, several proposals have been suggested to detect the non-existence of Cauchy horizon in some future observations \cite{Cao:2023par,Liu:2024iec,Wang:2025hla,Huang:2025gia,Guo:2025mwp,Lenzi:2025man}.

Nevertheless, the violation of the weak energy condition (WEC) as one of the conditions for the no-hair theorem by some nonpositive-definite scalar potentials $V(\phi)$ which minimally coupled with Einstein gravity could give rise to a class of asymptotically flat HBHs in the Einstein-Klein-Gordon (EKG) theory with a real scalar field $\phi$ \cite{Corichi:2005pa, Chew:2022enh, Chew:2024rin,Gubser:2005ih, Chew:2023olq,Chew:2024evh}. In particular, the inverted Higgs-like potential $V(\phi)=-\Lambda \phi^4 + \mu \phi^2$ is negative when $|\phi| > \sqrt{\mu/\Lambda}$, thus exterior solutions of HBHs can be emerged from the Schwarzschild black hole when $\phi$ is nontrivial at the event horizon \cite{Gubser:2005ih,Chew:2023olq}. Recently, its shadow has been studied using the ray-tracing approach \cite{Lim:2025cne}, where
the differences between the Schwarzschild black hole and HBHs with both having the same horizon radius, and found that the size of the shadow increases as $\phi_H$ increases, but the brightness of the rings remain nearly unaffected, this implies our HBHs can potentially mimic the Schwarzschild black hole if we vary the horizon radius of the HBHs. However, their interior structures remain largely unexplored.  

Therefore, our paper aims to perform a systematic, theory-spanning investigation of HBH interiors, focusing on whether the destruction of the Cauchy horizon is also a generic feature for this class of HBHs with the violation of WEC. The findings will have direct implications for the SCCC and will advance our understanding of black hole interiors in theories with nontrivial matter content. 

The structure of our paper is organized as follows. In Sec.~\ref{sec:th}, we briefly present the numerical construction of our HBH model, including the EKG theory, the Ansatz of metric, the set of coupled differential equations and the asymptotic behaviour of the functions near the event horizon. In Sec~\ref{sec:geo}, we briefly review the properties of exterior properties of HBHs from our previous work \cite{Chew:2023olq}. In Sec~\ref{sec:res}, we present and discuss our numerical findings on the interior of HBHs. Finally, we summarize our work and present an outlook in Sec.~\ref{sec:con}. 

\section{Theoretical Setting} \label{sec:th}


The Einstein gravity minimally coupled with a scalar potential $V(\phi)$ of a scalar field $\phi$ in the EKG theory:
\begin{equation} \label{EHaction}
 S=  \int d^4 x \sqrt{-g}  \left[  \frac{R}{16 \pi G} - \frac{1}{2} \nabla_\mu \phi \nabla^\mu \phi - V(\phi) \right]  \,.
\end{equation}
The variation of Eq.~\eqref{EHaction} with respect to $g_{\mu \nu}$ and $\phi$ yields the Einstein and Klein-Gordon equations, respectively
\begin{equation} 
 R_{\mu \nu} - \frac{1}{2} g_{\mu \nu} R =  8\pi G  \left(  \nabla_\mu \phi \nabla_\nu \phi -    \frac{1}{2} g_{\mu \nu} \nabla_\alpha \phi \nabla^\alpha \phi - g_{\mu \nu}  V(\phi)  \right) \,,  \quad  \nabla_\mu \nabla^\mu \phi  =  \frac{d V}{d \phi} \,.  
\end{equation}

The spacetime of static and spherically symmetric HBHs can be described by the following Ansatz:
\begin{equation}  \label{line_element}
ds^2 = - N(r) e^{-2 \sigma(r)} dt^2 + \frac{ dr^2}{N(r)} + r^2  \left( d \theta^2+\sin^2 \theta d\varphi^2 \right) \,, 
\end{equation}
where $N(r)=1-2m(r)/r$ with $m(r)$ is the Misner-Sharp mass function. The mass of HBH can be read off in the far asymptotic limit as $m(\infty)=M$, which $M$ is the Arnowitt-Deser-Misner (ADM) mass.

The substitution of Eq.\eqref{line_element} into the equations of motion yields a set of nonlinear ODEs:
\begin{equation}
m' = 4\pi G r^2 \left( \frac{1}{2} N \phi'^2 + V \right) \,, \quad \sigma' = - 4 \pi G r \phi'^2 \,,    \quad
\left(  e^{- \sigma} r^2 N \phi' \right)' = e^{- \sigma} r^2  \frac{d V}{d \phi} \,, \label{odes}
\end{equation}
where the prime $(')$ denotes the derivative of the functions with respect to the radial coordinate $r$.

In this paper, we adopt the following $V(\phi)$ to construct the HBH, which bifurcated from the Schwarzschild black hole when $\phi$ is nontrivial at the event horizon \cite{Chew:2023olq}:
\begin{equation} \label{vpot}
 V(\phi) = -\Lambda \phi^4 + \mu \phi^2 \,,
\end{equation}
where $\Lambda$, $\mu$ are the real-valued constants. $V(\phi)$ is unbounded from below, but possesses the false vacuum at $\phi=0$ and two degenerate maxima at $\phi_\text{max}=\pm \sqrt{\mu/(2\Lambda)}$. The mass of scalar field is $\sqrt{2 \mu}$. The solutions of self-gravitating scalaron can be constructed numerically by adopting Eq.~\eqref{vpot}, which demonstrates that they can be smoothly connected with the HBH in the small horizon limit \cite{Chew:2024bec}.

\section{The Exterior Solutions of HBHs} \label{sec:geo}

The exterior of globally regular solutions of HBHs obtained by directly integrating the ODEs numerically from the horizon $r_H$ to infinity by using two ODE solver packages: Colsys \cite{Ascher:1979iha} and Matlab bvp4c \cite{kierzenka2001bvp}. Colsys adopts the Newton-Raphson method to tackle the boundary value problems for the nonlinear ODEs, incorporating adaptive mesh refinement to produce solutions with more than 1000 points to ensure high precision and provide estimation of error. Besides, bvp4c is a boundary value problem solver with adaptive meshing, based on the three-stage Lobatto IIa collocation method.

\begin{figure}
\centering
\mbox{
\includegraphics[angle =0,scale=0.58]{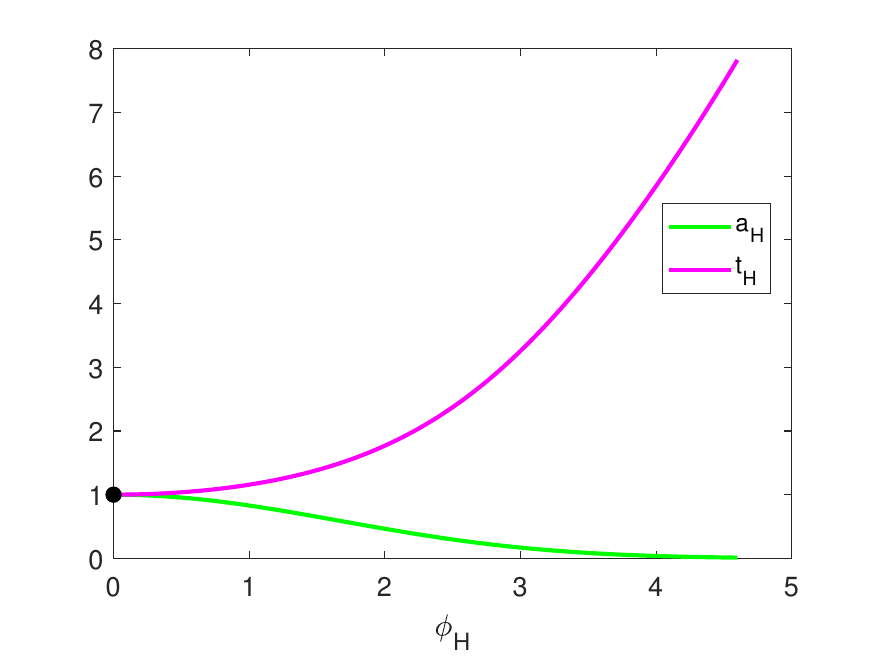}
}
\caption{Two basic properties of HBHs as the function of $\phi_H$: reduced area of horizon $a_H$ (purple) and reduced Hawking temperature $t_H$ (green). Black dot represents the Schwarzschild black hole.}
\label{aHtH}
\end{figure}

For the numerical calculations, the compactified radial coordinate $x=1-r_H/r$ is employed where $x$ ranges over $[0,1]$ with each end corresponding to the horizon and spatial infinity. At the horizon, the functions admit power series expansions with the following leading terms:
\begin{align}
 m(r) &= \frac{r_H}{2}+ 4 \pi G r^2_H (r-r_H) + O\left( (r-r_H)^2 \right) \,, \label{b1} \\
\sigma(r) &= \sigma_H  -  4 \pi G r_H \phi_1^2   (r-r_H) + O\left( (r-r_H)^2 \right)  \,, \\
 \phi(r) &= \phi_H +  \phi_1 (r-r_H) + O\left( (r-r_H)^2 \right)  \,, \label{b3}
\end{align} 
where
\begin{equation}
\phi_1= \frac{r_H \frac{d V(\phi_H)}{d \phi}}{1-8 \pi G r_H^2 V(\phi_H)}  \,,
\end{equation}
with $\sigma_H$ and $\phi_H$ denote the horizon values of $\sigma(r)$ and $\phi(r)$, respectively. In order to ensure finiteness of $\sigma(r)$ and $\phi(r)$ at $r = r_H$, the denominator in $\phi_1$ must obey $1 - 8 \pi G r_H^2 V(\phi_H) \neq 0$. At infinity, the asymptotic flatness requires $m(\infty) = M$ and $\sigma(\infty) = \phi(\infty) = 0$. Among the free parameters $\sigma_H$, $\phi_H$, $M$, $\Lambda$, $r_H$, and $\mu$, the values $\sigma_H$, $\phi_H$ and $M$ are fixed precisely by the boundary conditions. To streamline the numerics, we rescale the parameters via $r \rightarrow r/\sqrt{\mu}$, $m \rightarrow  m/\sqrt{\mu}$, $\phi \rightarrow \phi/\sqrt{8 \pi G}$, $\Lambda \rightarrow 8 \pi G \Lambda \mu$. This reduces the system to just two governing parameters: $r_H$ and $\Lambda$ for the HBH solutions.

Black holes typically described by the area of horizon $A_H$ and Hawking temperature $T_H$,  
\begin{equation}
 A_H = 4 \pi r^2_H \,, \quad  T_H = \frac{1}{4 \pi} N'(r_H) e^{-\sigma_H}  \,.
\end{equation}
The deviation of the HBHs from the Schwarzschild black hole can be quantified by two ``reduced" quantities at the horizon, which are $a_H$ is the reduced area of horizon and $t_H$ is the reduced Hawking temperature,
\begin{equation}
 a_H = \frac{A_H}{16 \pi M^2} \,, \quad t_H = 8 \pi T_H M \,.
\end{equation}
The Schwarzschild black hole is the trivial solution to the ODEs when $\phi_H=0$ at the horizon. However, the HBHs bifurcated from the Schwarzschild black hole when $\phi_H$ is nontrivial at the horizon. Recall that $a_H=t_H=1$ for the Schwarzschild black hole with $r_H=2M$ and $\sigma_H=0$ when $\phi_H=0$, as represented by the black dot in Fig.~\ref{aHtH}. Note that $a_H$ always equal to unity for the Schwarzschild black hole regardless of horizon size. As $\phi_H$ increases from zero, $a_H$ decreases from 1 and eventually approaches zero. This implies that $M$ increases as $a_H$ decreases, since $a_H$ is inversely proportional to $M$. When $\phi_H$ becomes larger, the scalar field contributes additional energy to the HBHs, which increases its mass $M$. This results in a smaller $a_H$, as the HBHs become more massive than the Schwarzschild black hole. In contrast, $t_H$ monotonically increases from 1 \cite{Chew:2023olq}.

\section{The Interior Solutions of HBHs}\label{sec:res}

\begin{figure}
\centering
\mbox{
(a)
\includegraphics[angle =0,scale=0.58]{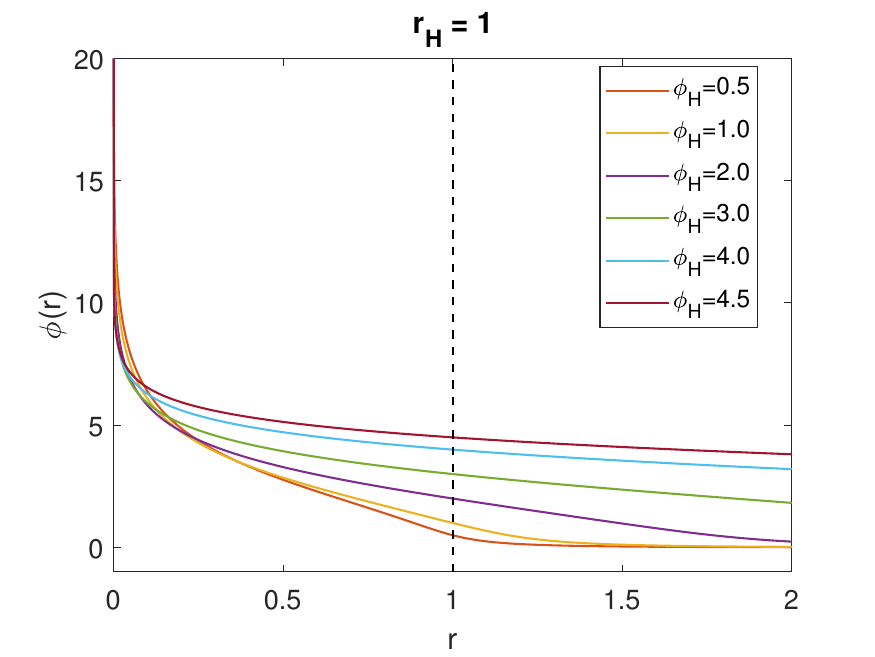}
(b)
\includegraphics[angle =0,scale=0.58]{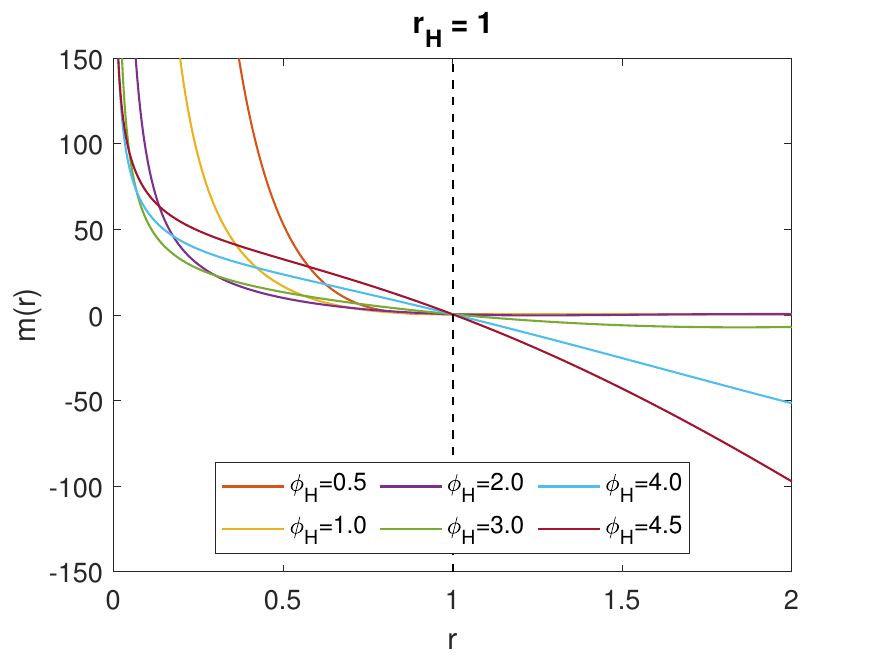}
}
\mbox{
(c)
\includegraphics[angle =0,scale=0.58]{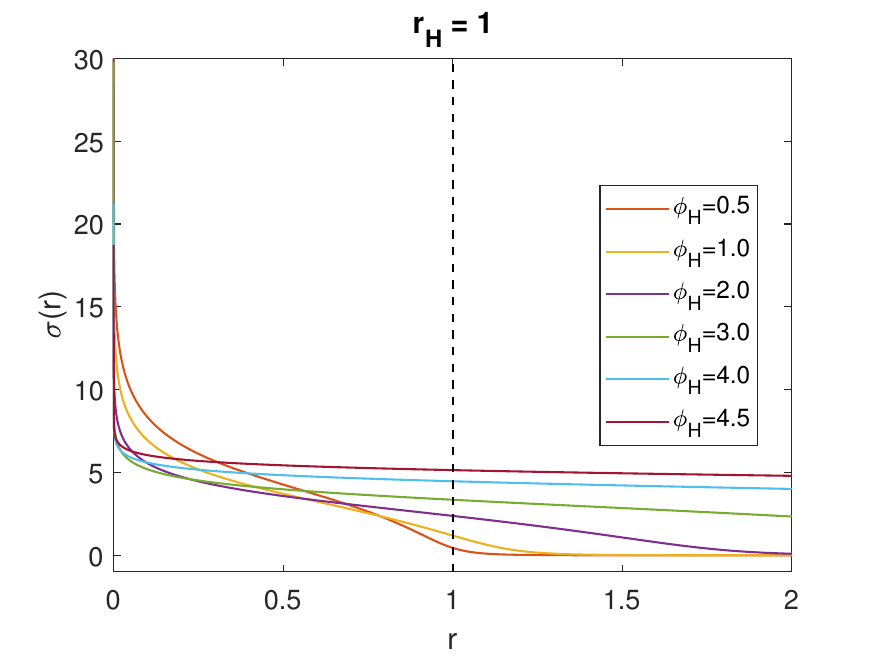}
(d)
\includegraphics[angle =0,scale=0.58]{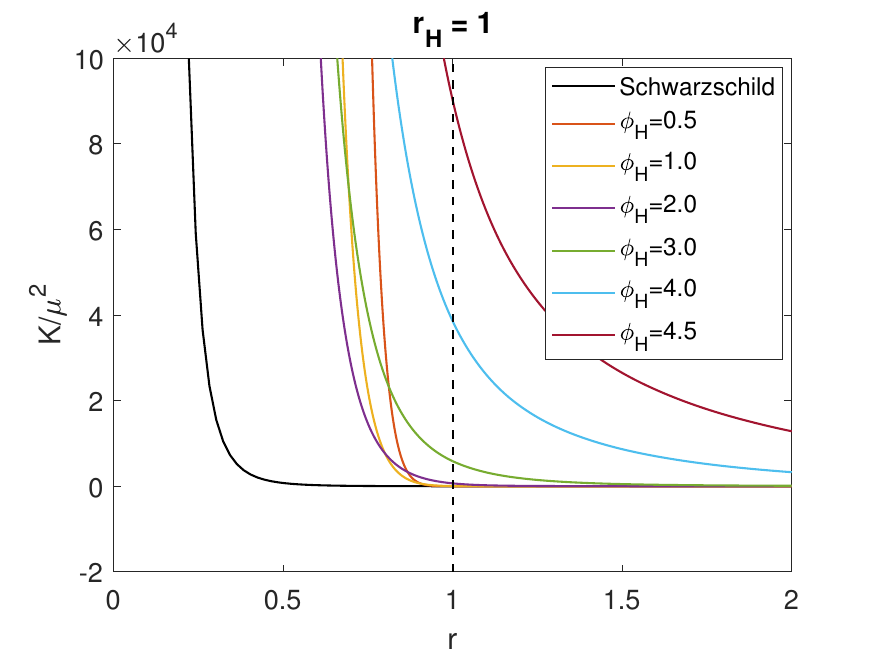}
}
\mbox{
(e)
\includegraphics[angle =0,scale=0.58]{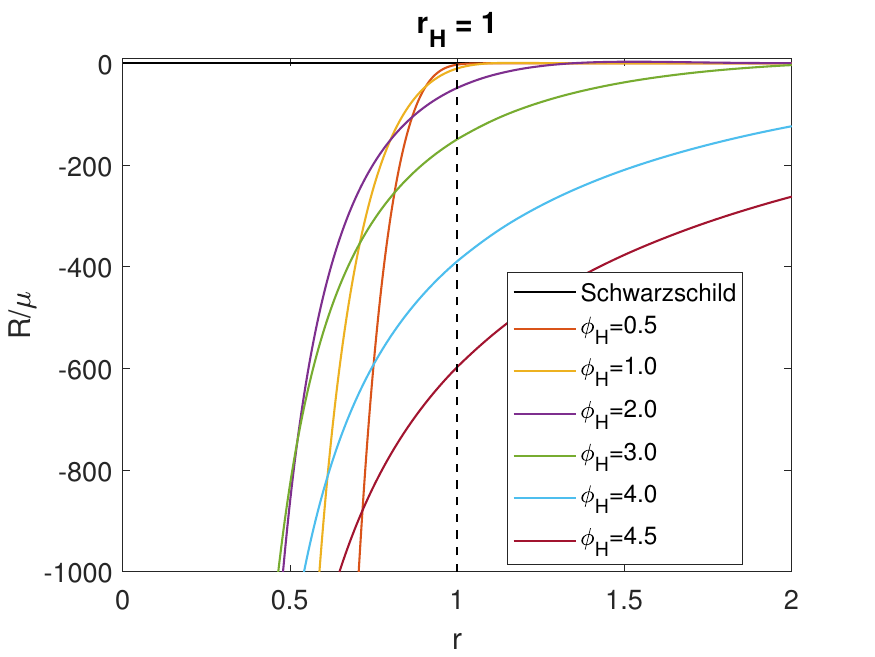}
(f)
\includegraphics[angle =0,scale=0.58]{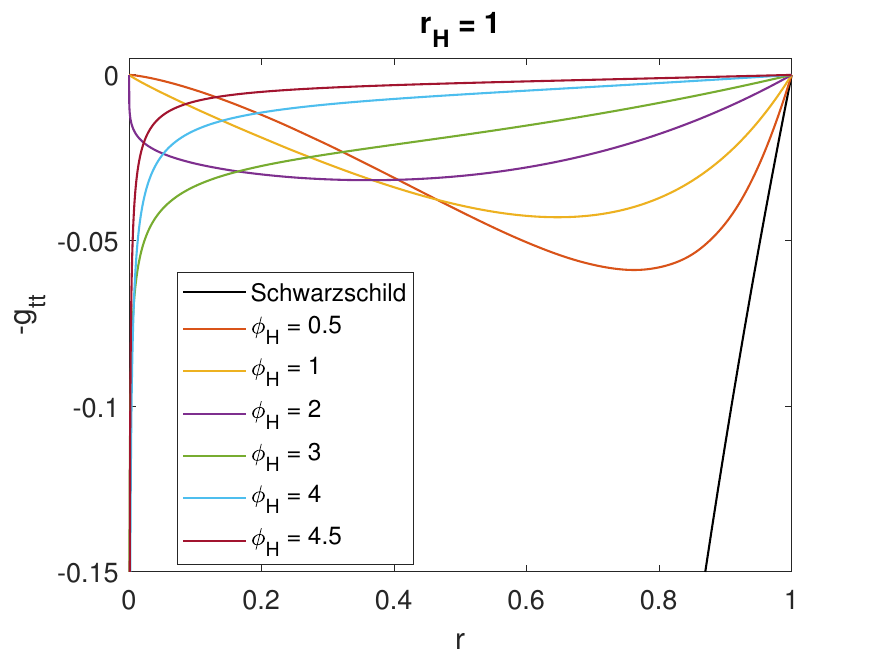}
}
\caption{Properties inside the HBHs $(r<r_H)$ with $r_H=1$ for several values of $\phi_H$: (a) $\phi(r)$; (b) $m(r)$; (c) $\sigma(r)$; (d) The scaled Kretschmann scalar $K/\mu^2$; (e) The scaled Ricci scalar $R/\mu$; (f) The metric component $-g_{tt}$.}
\label{sol_rH_1}
\end{figure}

The interior solutions of HBHs are obtained by directly integrating Eq.~\eqref{odes} inward from the horizon $r_H$, imposing the same boundary conditions (Eqs.~\eqref{b1}–\eqref{b3}) at $r_H$. In Fig.~\ref{sol_rH_1}(a), the profiles of $\phi(r)$ for several values of $\phi_H$ inside the horizon ($r<r_H$) increase monotonically and then diverge as $r \rightarrow 0$. The growth of $\phi(r)$ becomes significantly steeper as $\phi_H$ increases. Figs.~\ref{sol_rH_1}(b) and (c) show that the metric functions $m(r)$ and $\sigma(r)$ exhibit the same qualitative behavior: both increase monotonically and diverge as $r \rightarrow 0$, reflecting their dependence on $\phi(r)$ and its derivative. Thus, all three functions display monotonic behavior and divergence near $r=0$, in contrast to the four distinct stages observed in holographic superconductors. Moreover, the divergence of the mass function $m(r)$ near $r=0$ suggests that mass inflation may occur in the vicinity of the singularity, which is presumably located at $r=0$.

To determine whether $r=0$ corresponds to a curvature singularity, we examine the Kretschmann scalar $K$ which reads
\begin{equation}
 K = R^{\alpha \beta \gamma \delta} R_{\alpha \beta \gamma \delta} = \left(  3 \sigma' N' + 2 N \sigma'' - N'' -2 N \sigma'^2  \right)^2    +  \frac{2}{r^2} \left( N'-2 N \sigma'  \right)^2 + \frac{2 N'^2}{r^2} + \frac{4 (N-1)^2}{r^4}\,. 
\end{equation}
The divergence of $K$ in the vicinity of $r \rightarrow 0$, indicating the presence of the curvature singularity. Then we could simplify the above expression using the ODEs and rescaling of parameters, 
\begin{align}
K/\mu^2
&= \frac{1}{r^4} \Bigg[
\left( r^2 N \phi'^2 - 2N + 2 \right)^2
+ \frac{1}{2} \left( - r^2 N \phi'^2 + 2 r^2 V + 2N - 2 \right)^2  \nonumber \\
&\qquad\qquad
+ \frac{1}{2} \left( r^2 N \phi'^2 + 2 r^2 V + 2N - 2 \right)^2
+ 4 (N-1)^2
\Bigg] \,,
\end{align}
where all quantities are understood to be dimensionless after rescaling. For comparison, in the Schwarzschild limit where the scalar field vanishes at the horizon and throughout the spacetime, one recovers $K=12 r_H^2/r^6$ which diverges as $r \rightarrow 0$, confirming that $r=0$ is a curvature singularity in the vacuum case. The numerical results for the HBHs are shown in Fig.~\ref{sol_rH_1}(d). The scaled quantity $K/\mu^2$ increases monotonically for $r<r_H$ and diverges as $r \rightarrow 0$, indicating the existence of the curvature singularity at the origin. Moreover, the growth of $K/\mu^2$ becomes increasingly steep as $\phi_H$ increases, suggesting that stronger scalar hair enhances the curvature blow-up in the deep interior. 

We further examine the Ricci scalar $R$ to probe the interior structure of the HBHs, where its explicit form is:
\begin{align}
     R = R^\alpha\,_\alpha &= -N'' + \frac{3 r \sigma'-4}{r} N' + \frac{2 \left( 2 r N \sigma' - N +1 + r^2 N\sigma'' - r^2 N \sigma'^2  \right)}{r^2} \,. \nonumber \\
  \Rightarrow R/\mu   &=   \phi'^2 N + 4 V \,.
\end{align}
The scaled $R/\mu$ is also obtained using the ODEs and rescaling of parameters. Recall that $R$ identically vanishes for the Schwarzschild black hole, reflecting its vacuum nature. In contrast, in the presence of a nontrivial scalar field, the spacetime is no longer Ricci-flat. From our previous analysis of the exterior geometry \cite{Chew:2023olq}, we found that $R$ is predominantly negative from spatial infinity down to the event horizon. Consistently, the scaled Ricci scalar $R/\mu$ shown in Fig.~\ref{sol_rH_1}(e) remains negative inside the horizon. As $r<r_H$, $R/\mu$ decreases monotonically and eventually diverges as $r \rightarrow 0$. This divergence signals the formation of a curvature singularity at $r=0$ and confirms that the scalar field significantly modifies the interior geometry compared to the Schwarzschild case.

Fig.~\ref{sol_rH_1}(f) reveals a peculiar feature in the behavior of $-g_{tt}=N(r)e^{-2 \sigma(r)}$ within the interior of HBHs with $r_H=1$. For the Schwarzschild case, $-g_{tt}$ decreases monotonically and diverges as $r \rightarrow 0$. In contrast, for HBHs with $\phi_H = 0.5, 1.0,$ and $2.0$, $-g_{tt}$ first decreases to a minimum and then increases toward zero as $r \rightarrow 0$, indicating that the singularity at $r=0$ is null-like. However, for larger values $\phi_H = 3.0, 4.0,$ and $4.6$, the behavior reverts to the Schwarzschild-like case: $-g_{tt}$ decreases monotonically and diverges as $r \rightarrow 0$, implying a spacelike singularity.

The presence of a Cauchy horizon would be indicated by an additional root of $N(r)$ for $r<r_H$. In our solutions, no such root is found, and therefore no Cauchy horizon appears inside the HBHs. This is consistent with our expectation, since the HBHs considered here are electrically neutral and bifurcate from the Schwarzschild solution (see Fig.~\ref{aHtH}), thereby inheriting its basic interior structure.

\begin{figure}
\centering
\mbox{
(a)
\includegraphics[angle =0,scale=0.58]{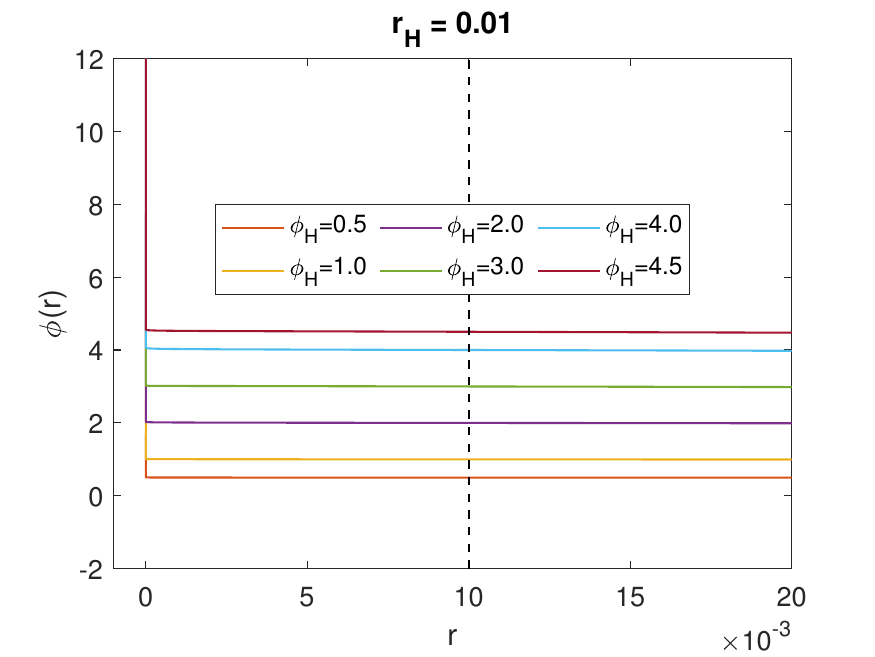}
(b)
\includegraphics[angle =0,scale=0.58]{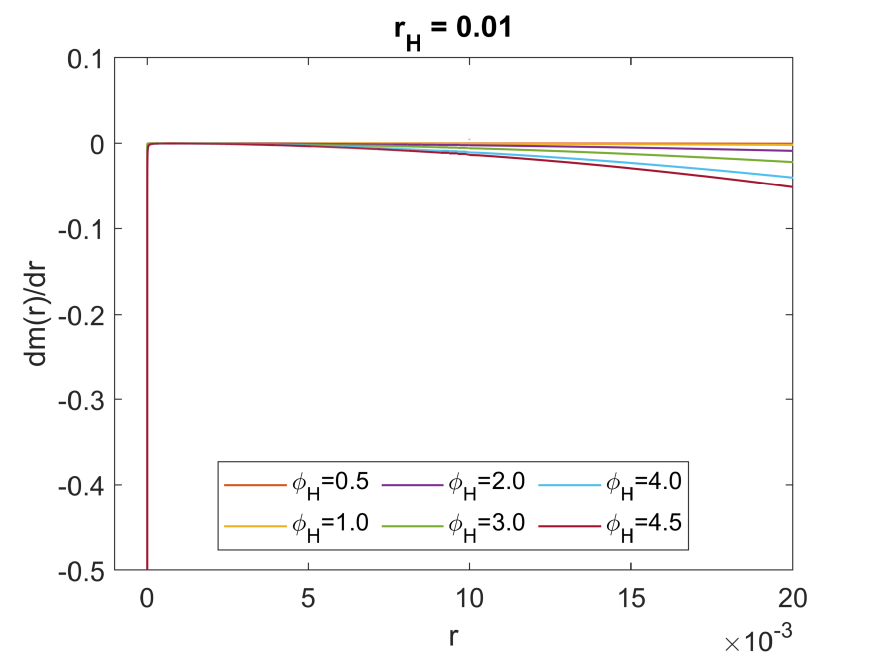}
}
\mbox{
(c)
\includegraphics[angle =0,scale=0.58]{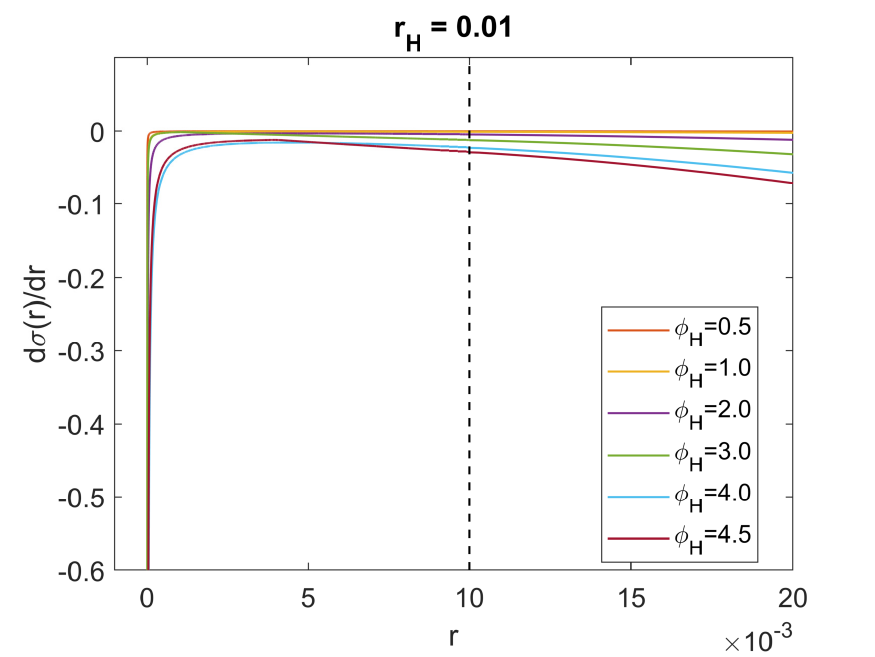}
(d)
\includegraphics[angle =0,scale=0.58]{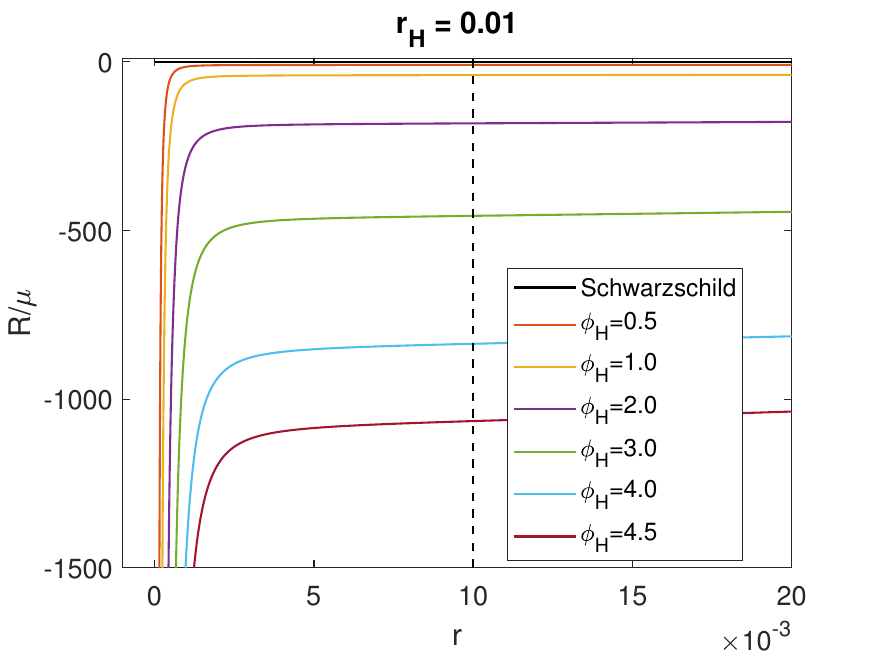}
}
\mbox{
(e)
\includegraphics[angle =0,scale=0.58]{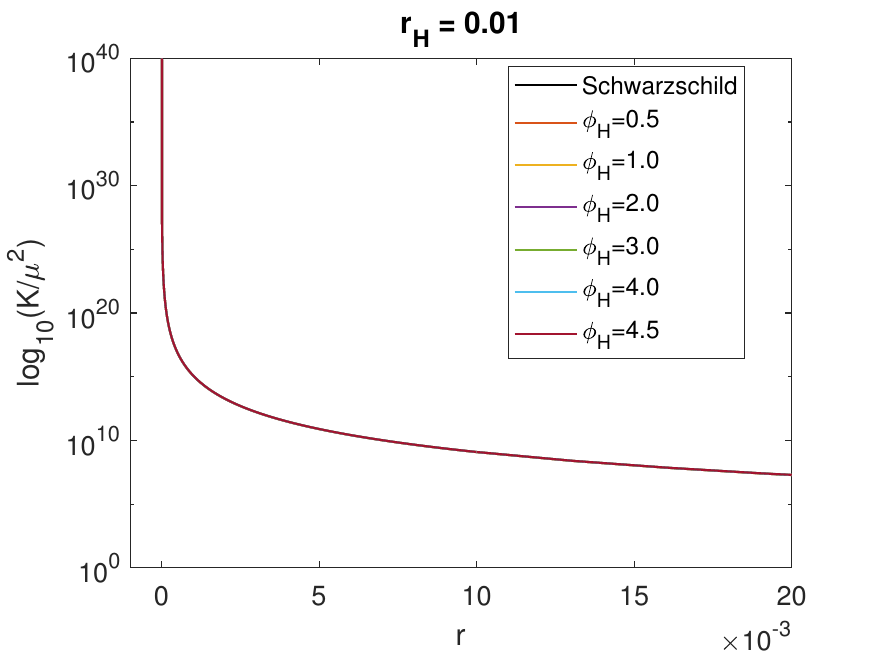}
(f)
\includegraphics[angle =0,scale=0.58]{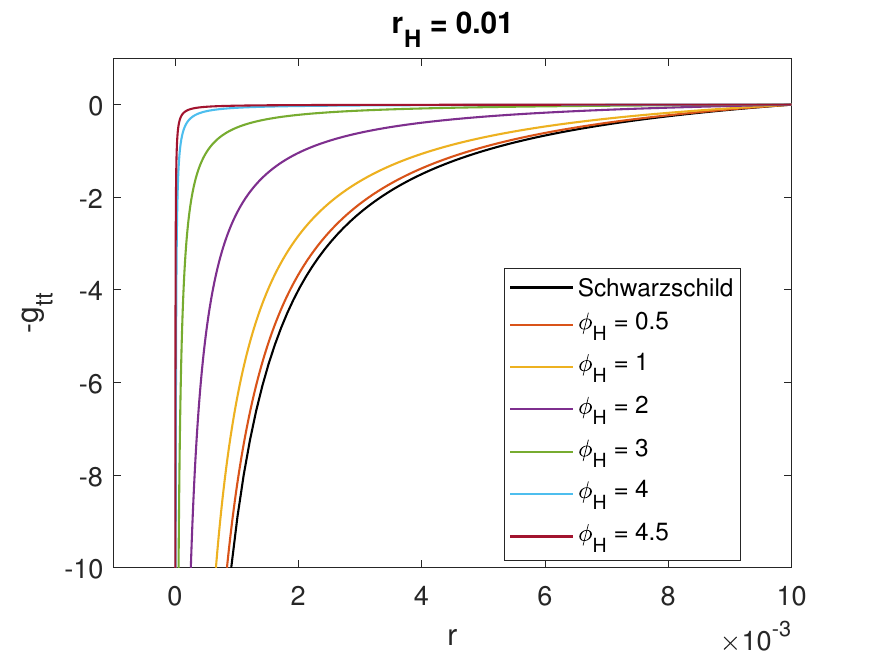}
}
\caption{Properties inside the HBHs with $r_H=0.01$ for several values of $\phi_H$: The profiles of functions: (a) $\phi(r)$; (b) $dm(r)/dr$; (c) $d\sigma(r)/dr$; (d) The scaled Ricci and (e) Kretschmann scalars.}
\label{sol_rH_0o01}
\end{figure}

As the horizon radius decreases to very small values, for example $r_H=0.01$ as shown in Fig.~\ref{sol_rH_0o01}, the scalar field $\phi(r)$ again increases monotonically and diverges as $r \rightarrow 0$, with steeper growth for larger $\phi_H$. The same behavior propagates to the metric functions $m(r)$ and $\sigma(r)$ which demonstrated by their derivatives in Figs.~\ref{sol_rH_0o01}(b) and (c), respectively, owing to their dependence on $\phi(r)$. Consequently, the scaled Ricci and Kretschmann scalars in Figs.~\ref{sol_rH_0o01}(d) and (e) diverge near $r=0$, confirming the presence of a curvature singularity whose strength increases with $\phi_H$. In this case, however, Fig.~\ref{sol_rH_0o01}(f) shows that $-g_{tt}$ decreases monotonically and diverges as $r \rightarrow 0$, indicating that the singularity remains spacelike. Although HBHs connect smoothly to self-gravitating scalaron solutions in the small-horizon limit, this transition is not directly visible in the interior, since the geometry still corresponds to that of an HBH.

\begin{figure}
\centering
\mbox{
(a)
\includegraphics[angle =0,scale=0.58]{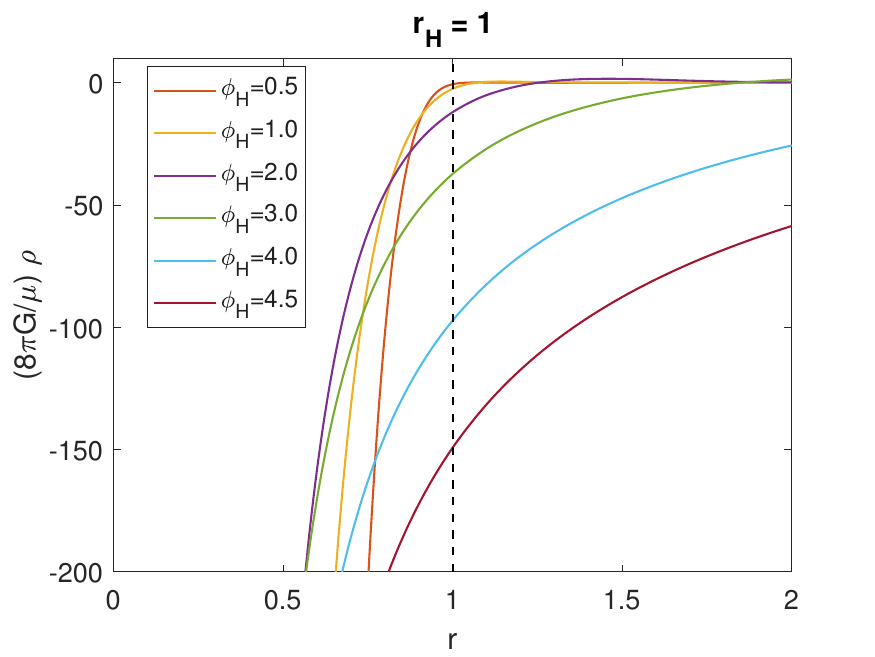}
(b)
\includegraphics[angle =0,scale=0.58]{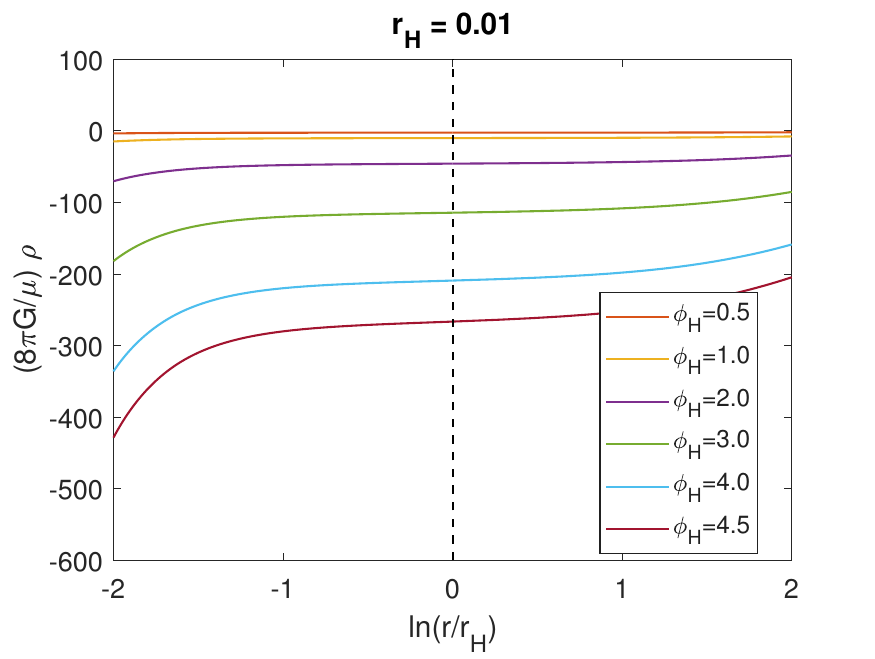}
}
\caption{The scaled weak energy condition $(8\pi G/\mu)\rho$ of HBHs with several $\phi_H$ for (a) $r_H=1.0$; (b) $r_H=0.01$.
}
\label{energy_cond}
\end{figure}

Because the existence of exterior HBH solutions that evade the no-hair theorem requires violation of the weak energy condition (WEC), defined by energy density $\rho=-T^{t}{}_{t}$, it is natural to examine whether the WEC is also violated inside the horizon. Fig.~\ref{energy_cond}(a) shows that the WEC remains violated inside HBHs with $r_H=1$: the scaled energy density $(8\pi G/\mu)\rho$ is negative, decreases monotonically for $r<r_H$, and diverges as $r \rightarrow 0$. This divergence is again associated with the divergence of $\phi(r)$ at the singularity, and the violation of the WEC becomes more severe as $\phi_H$ increases. Similar effect can also be observed for HBHs with $r_H=0.01$ in Fig.~\ref{energy_cond}(b).

\begin{figure}
\centering
\vskip-10pt
\mbox{
\includegraphics[angle =0,scale=0.85]{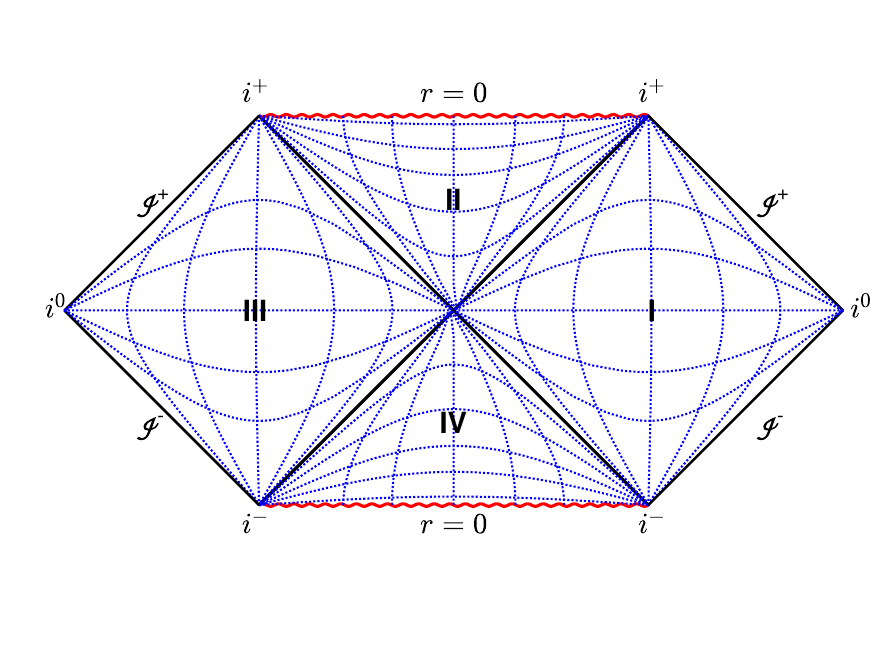}
}
\vskip-50pt
\caption{The Penrose diagram for HBH with $\phi_H=1$ for $r_H=1$, where its causal structure divided into four regions (I, II, III, IV), $i^{\pm}$ denotes the future $(+)$ and past $(-)$ timelike infinities, $i^0$ denotes the spatial infinity, $\mathcal{I}^\pm$ denotes the future $(+)$ and past $(-)$ null infinities, the red curvy line represents the spacelike singularity at $r=0$.} 
\label{Penrose}
\end{figure}

Finally, we present the causal structure of the HBHs using the Penrose diagram in Fig.~\ref{Penrose}. We maximally extend the spacetime of these solutions to remove the coordinate singularity at the horizon by changing the coordinate of metric (Eq.~\ref{line_element}) in the Kruskal-Szekeres (KS) coordinates $(u,v)$: $u = t - r_*$ and $v = t + r_*$, where the tortoise coordinate $r_*$ is defined as: $r_* = \int e^{\sigma(r)}/N(r) dr$. The Penrose diagram of the HBH divided by the horizon $r_H$ into four regions, which are I, II, III, IV using the following coordinate transformations: Region I ($U = -e^{-\kappa u}$, $V = e^{\kappa v}$) corresponds to the exterior part of spacetime with $r>r_H$, where an observer can escape freely to infinity; Region II ($U = e^{-\kappa u}$, $V = e^{\kappa v}$) corresponds to the interior part of spacetime with $r<r_H$, where an observer cannot escape to outside once crossing in and will reach to future singularity $i^+$; Region III ($U = -e^{-\kappa u}$, $V = e^{\kappa v}$) corresponds to another parallel universe as Region I; Region IV ($U = e^{-\kappa u}$, $V = e^{\kappa v}$) corresponds to the white hole region with a time-reversal direction, which connected to the past singularity $i^-$. In these coordinates $(U,V)$ the metric becomes regular at $r=r_H$ and takes the following form
\begin{equation}
ds^2 = -\frac{1}{\kappa^2} N(r) e^{-2 \sigma(r)} e^{-2\kappa r_*}  dU dV + r^2(U,V) \left( d\theta^2 + \sin^2 \theta d\varphi^2  \right) \,,
\end{equation}
where the surface gravity $\kappa$ is defined as $\kappa=2\pi T_H$, $r(U,V)$ is determined implicitly through the definition of $r_*$. The infinity of KS coordinates $(U,V)$ mapped to the compactified coordinates $(T,X)$ using the conformal transformation: $T = \frac{1}{2} (\arctan(V) + \arctan(U))$ and $X = \frac{1}{2} (\arctan(V) - \arctan(U))$.

In general, the causal structure of HBHs is qualitatively similar to that of a maximally extended Schwarzschild spacetime: a spacelike curvature singularity is hidden behind an event horizon, and no Cauchy horizon is present. However, in the special case where $-g_{tt}\rightarrow 0$ near $r=0$, the singularity can become null-like. Overall, the presence of the scalar field with an inverted Higgs potential preserves the globally hyperbolic character of the interior, and the singularity remains the inevitable future boundary for all observers inside the event horizon.

\section{Conclusion}\label{sec:con}

In this work, we have numerically obtained the interior solutions of HBHs by directly integrating the field equations inward from the event horizon. This provides an important step toward understanding the global structure of these solutions. We find that the scalar field and the metric functions exhibit monotonic increasing behavior inside the horizon and diverge as $r \rightarrow 0$. The divergence of the Ricci and Kretschmann scalars confirms that $r=0$ corresponds to a genuine curvature singularity. The mass inflation also occurs in the vicinity of the singularity. We also observe that the weak energy condition (WEC) is violated throughout the interior region, and the degree of violation becomes more pronounced as the scalar field at the horizon increases.

The behavior of the metric component $-g_{tt}$ reveals additional structure in the nature of the singularity. In most cases, $-g_{tt}$ decreases monotonically and diverges as $r \rightarrow 0$, indicating a spacelike singularity. However, for certain parameter ranges, $-g_{tt}$ decreases to a minimum and then increases toward zero, suggesting that the singularity can become null-like. The analysis of the metric function $N(r)$ also allows us to probe the possible existence of a Cauchy horizon. For the electrically neutral HBHs studied here, no additional root of $N(r)$ is found inside the horizon, and therefore no Cauchy horizon forms.

This naturally raises the question of whether a Cauchy horizon could appear in electrically charged hairy black holes, or whether the presence of scalar hair may suppress its formation. Since Cauchy horizons are known to arise in charged black holes, it would be interesting to investigate whether similar structures persist in charged HBH solutions, such as those supported by asymmetric scalar potential \cite{Chew:2024rin}. We leave this problem for future work.

It is also important to note that the interior structure of black holes in realistic gravitational collapse is generally dynamical rather than static. Previous studies using the double-null formalism have shown that the behavior of the Cauchy horizon during gravitational collapse can differ significantly from that inferred from static solutions \cite{Chew:2023upu}. For example, in certain scalar–tensor theories, the Cauchy horizon may fail to form during collapse, revealing that a scalar field can strongly influence the internal causal structure. These results provide further motivation to construct and analyze charged hairy black holes as a first step toward studying their dynamical evolution.

Another motivation for studying the interior of HBHs comes from quantum gravity. The black-hole interior can often be modeled by a Kantowski–Sachs metric \cite{Kantowski:1966te,Collins:1977fg,Doran:2006dq,Bouhmadi-Lopez:2020wve}, which provides a useful framework for formulating the Wheeler–DeWitt equation describing the quantum state of the interior  \cite{Bouhmadi-Lopez:2019kkt}. An interesting open question is how the presence of scalar hair modifies this equation and whether quantum effects could resolve or soften the classical singularity \cite{Chien:2025tzm}. Although such issues lie beyond the scope of the present work, the detailed classical solutions obtained here provide an essential foundation for future investigations of the quantum dynamics of HBHs interiors.

\section*{Acknowledgement}
 XYC is supported by the starting grant of Jiangsu University of Science and Technology (JUST) and National Science Foundation of China (no: W2533026). XYC would like to acknowledge the hospitality provided by Jutta Kunz during the visit to the University of Oldenburg, by the organizer of Persidangan Fizik Kebangsaan (PERFIK 2025), by Yen-Kheng Lim at Xiamen University Malaysia, and by the gravity group at Jiangxi Normal University. DY was supported by the National Research Foundation of Korea (Grant No.: 2021R1C1C1008622, 2021R1A4A5031460). We are grateful to have useful discussion with Yen Chin Ong.


\bibliography{mybiblio}



\end{document}